\documentclass[aps,pre,amsfonts,amssymb,amsmath,floatfix,twocolumn,
showpacs]{revtex4}

\usepackage{graphicx}

\begin{document}     

\title{Comment on ``Parameter-free scaling for 
nonequilibrium growth processes"}

\author{A. Kolakowska}
\affiliation{Department of Physics, 
The University of Memphis, Memphis, TN 38152} 
\date{\today}

\begin{abstract} 
In the paper [Phys.~Rev.~E \textbf{79}, 051605 (2009)] 
by Chou and Pleimling a claim is made that 
a parameter-free scaling that gives  
data collapse for some simulation models 
would replace universal Family-Vicsek (FV) scaling. Here, 
by giving the explicit form of this scaling for 
competitive growth models, it is shown that 
data collapse in this procedure is obtained 
by a shift-and-scale operator that gives no  
information about stochastic 
dynamics and has no relation with FV function.

\end{abstract}


\maketitle

\noindent
Nonequilibrium sufrace-growth processes of  
Ref.~\cite{CP09} are SOS models 
in $(1+1)$ dimensions with periodic boundary and  
time $t$ is a number of deposited monolayers 
to a substrate of $L$ sites.  
At $t=0$ the substrate is flat.  
The growth rule is a competitive growth model:  
``\textit{either} RD (active with probability $q$) 
\textit{or} X (active with probability $p=1-q$)," 
where RD is random deposition and X is a process that builds 
correlations. For these models, 
the initial time evolution of the surface width $w(t)$ has 
two growth regimes before cross-over time to saturation 
\cite{CP09,many}.

In processes ``RD or X" 
for any $L$ and $p$, $w(t)$   
has three evolution regimes, as shown in Fig.1 
of Ref.\cite{CP09}:
\begin{equation}
\label{eq01} 
t \in [0;t_1] \cup (t_1;t_2) \cup [t_2;+ \infty)=[0;+\infty).
\end{equation}
First, $w(t)$ obeys the RD 
universal power law:
\begin{equation}
\label{eq02} 
\forall t \in [0;t_1]: w(t) \propto t^{\beta _1}, \quad \beta_1=1/2.
\end{equation}
Note, $\beta_1$ 
is \textit{not} a scaling exponent \cite{note01}. 
Later, $w(t)$ obeys: 
\begin{eqnarray}
\forall t \in (t_1;t_2) &:&  w(t) \propto t^{\beta_2}, \label{eq03} \\ 
\forall t \in [t_2;+\infty) &:&  w(t) \propto L^{\alpha_2}, \label{eq04} 
\end{eqnarray}
where $\beta_2$ and $\alpha_2$ are  
\textit{scaling} exponents (growth and roughness, 
respectively) of the universality class of X.

For fixed $L$ and $p$, values of $w(t)$ are in the interval 
$[0;w_2]$, where $w_2=w(t_2) \propto L^{\alpha_2}$ by  
Eq.(\ref{eq04}). After Ref.\cite{CP09}, this interval is  
$[0;w_2]= [0;w_1] \cup (w_1;w_2],$
where $w_1=w(t_1) \propto \sqrt{t_1}$ by Eq.(\ref{eq02}). 
In the limit of large but finite $L$, 
$w_2$ is large but finite; hence, in general:
\begin{equation}
\label{eq06}
w(t) \in [0;w_1] \cup (w_1;w_2] = [0;w_2].
\end{equation}

In Eqs.(\ref{eq01}) and (\ref{eq06}), $t_2$ and 
$w_2$ are functions of $L$. But, $w(t)$ 
in Eq.(\ref{eq02}) does \textit{not} depend on $L$ \cite{note01}.  
In Ref.\cite{CP09} this is seen in Fig.2a, 
where $w(t)$ is for the model 
``RD or X" and X is `random deposition 
with surface relaxation,' i.e.,   
in Edwards-Wilkinson (EW) universality class. 
For the model ``RD or EW," the data 
can be summarized as the family of curves  
parameterized by $p$ and $L$:
\begin{equation}
\label{eq07}
w(t;L;p)=\left\{ \begin{array}
{r@ {\quad , \quad} l}
c_1 \sqrt{t} & t\in [0;t_1(p)] \\
c_2 t^{\beta_2} & t\in (t_1(p);t_2(L,p)) \\
c_3 L^{\alpha_2} & t\in [t_2(L,p);+\infty]  ,
\end{array} \right. 
\end{equation}
where $c_1$, $c_2$, and $c_3$ are constants.

In the language of `data collapse,' Eq.(\ref{eq07}) says that 
for $t\in [0;t_1(p)]$ \textit{all} the data for \textit{all} $L$ 
and for \textit{all} $p$ are \textit{already} in  
one curve $w=c_1 \sqrt{t}$. In order to collapse 
the data for $t\in (t_1(p);+\infty)$ it is required 
to \textit{simultaneously} multiply $t$ 
and $w(t)$ by \textit{different} scale factors 
$a(p,L)$ and $b(p,L)$:
\begin{equation}
\label{eq08}
t \to t/a(p,L), \quad \mathrm{and} \quad w(t) \to w(t)/b(p,L).
\end{equation}
But, if affine transformations in Eq.(\ref{eq08}) 
give data collapse for $t\in (t_1(p);+\infty)$, when 
they are applied to $w(t)=c_1\sqrt{t}$ for $t\in [0;t_1(p)]$ 
they will produce `data scatter.' This is because 
$a(p,L) \ne b(p,L)$ and the data for $t\in [0;t_1(p)]$ 
already follow one curve for any $L$ and $p$. 
That is, Family-Vicsek (FV) scaling produces data 
collapse in EW-scaling regime for 
$t\in (t_1;+\infty)$ and destroys data coalescence 
in RD-growth regime for $t\in [0;t_1]$.

Because of this, the questions are about the 
rationale for the data collapse shown  
in Ref.\cite{CP09} and about a relevance of the proposed parameter-free 
scaling to the universal dynamic scaling. The answers are given below.

In Ref.\cite{CP09}, data are collapsed in two-step transformations  
performed individually for each curve of Eq.(\ref{eq07}), 
i.e., for each set of points $(t,w)$ representing one curve  
indexed by $p$ and $L$. 
In the first step, $t$ is divided by $t_1$, 
and $w(t)$ is divided by $w_1$, and the log 
is taken of all numbers. In this way intervals in 
Eq.(\ref{eq01}) are mapped onto 
$t \to t'' \in (-\infty ;0] \cup (0; \tau'')  
\cup [\tau'' ;+ \infty)=(-\infty;+\infty)$,  
where $\tau''=\log{(t_2/t_1)}$. The intervals in  
Eq.(\ref{eq06}) are mapped onto 
$w \to w'' \in (-\infty ;0] \cup (0; w_2''] =  
(-\infty; w_2''] $, 
where $w_2''=\log{(w_2/w_1)}$. 
In the second step, a number $\lambda$ is selected such that 
$\lambda w_2''=1$, and $t''$ and $w''$ are 
multiplied by $\lambda$. This gives 
\begin{eqnarray}
w'' &\to& w' \in (-\infty ;0] \cup (0;1] = (-\infty; 1] , \label{eq12} \\ 
t'' &\to& t' \in (-\infty ;0] \cup (0;\tau) \cup [\tau; +\infty), \label{eq13}
\end{eqnarray}
where $\tau = \tau''/w_2''=1/\beta$. In effect, 
original $w$ values in Eq.(\ref{eq06}) are mapped onto Eq.(\ref{eq12}), 
and $t$ values are mapped from Eq.(\ref{eq01}) onto Eq.(\ref{eq13}). 
The outcome of this mapping is one curve, regardless of 
the pair $p$ and $L$. After transformation, RD-growth phase is in 
$(-\infty;0] \times (-\infty;0]$, correlation-growth phase is in 
$(0;\tau) \times (0;1)$, and saturation-growth phase is in 
$[\tau;+\infty ) \times \{1 \}$.

For curves in Eq.(\ref{eq07}) (shown here in the left Fig.1) the 
procedure of Ref.\cite{CP09} is described by the operator 
$\hat{G}_{(p,L)}$,
\begin{equation}
\label{eqx01}
\hat{G}_{(p,L)}: (x,y) \longrightarrow (x',y'),
\end{equation}
where $x=\log{t}$, $y=\log{w}$, $x'=\log{t'}$, 
and $y'=\log{w'}$. Denoting $x_i=\log{t_i}$ and  
$y_i=\log{w_i}$ for $i=1,2$, standard algebra methods  
give the explicit form of $\hat{G}_{(p,L)}$: 
\begin{equation}
\label{eqx02}
x \to x' = \frac{x-x_1}{x_2-x_1} \frac{1}{\beta_2} \: , \; \; \;
y \to y' = \frac{y-y_1}{y_2-y_1} \, .
\end{equation}
Explicitly, $\hat{G}_{(p,L)}=\hat{S}_{(p,L)} \circ \hat{T}_{(p)}$ 
is the composition of the translation operator,  
$\hat{T}_{(p)} : (x,y) \to (x'',y'') = (x-x_1,y-y_1)$, 
and the scale operator, 
$\hat{S}_{(p,L)} : (x'',y'') \to (x',y') =        
( \frac{x''/\beta_2}{x_2-x_1},\frac{y''}{y_2-y_1})$, 
in $(x,y)$-plane. Thus, 
$\hat{G}_{(p,L)}$ is a shift-and-scale 
operator that translates the curves  
to one position and adjusts 
the length of the correlation-growth phase to 
$\sqrt{\tau^2+1}$, as shown here in Fig.1. 
Such shift-and-scale operation is possible 
because each curve in this family carries one universal 
footprint: the initial RD transient 
--- where each curve has the same one slope of $1/2$  
but its length depends on $p$ ---  
that is followed by a specific universal correlation phase, 
where each curve has the same one slope of $\beta_2$ 
and ends at saturation phase. 
However, a similar picture may or may not occur 
in other competitive growth phenomena, so 
$\hat{G}_{(p,L)}$ is not general. Moreover, 
as this explicit derivation of $\hat{G}_{(p,L)}$ shows, 
the shift-and-scale operator in $(x,y)$-plane is not 
a dynamic scaling.

\begin{figure}[t]
\includegraphics[width=8.0cm]{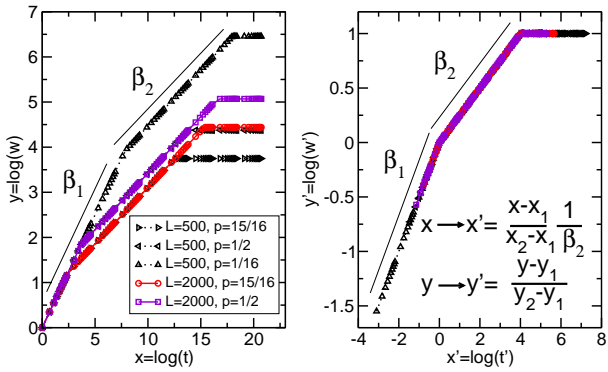}
\caption{\label{fig1} (Color on line)
Domain (left) and image (right) of the shift-and-scale operator
$\hat{G}_{(p,L)}$. In the image, cross-over to saturation is at
$(\tau =1/\beta,1)$. The curves are for ``RD or EW''
competitive growth process in Eq.(\ref{eq07}).
Here, $\log(\cdot) \equiv \log_{10} (\cdot)$.
}
\end{figure}

Data collapse by shift-and-scale operation 
is a nice illustration of the  
known fact that all systems in one universality 
class follow one universal curve. 
But it has no connection with  
finite-size dynamic scaling and with FV function. 
Affine scaling of interfaces such as the 
one in Eq.(\ref{eq08}) reflects universal 
dynamics of correlations. FV function summarizes 
a relation between scaling properties of 
growing surfaces and symmetry properties of 
equations that describe growth dynamics. 
In FV function, argument and 
prefactor contain \textit{explicit} information about 
the way the dynamics is affected by growth parameters. 
Physics wise, FV function 
provides explanation not only for data 
collapse but first of all for the universal shape of $w(t)$, 
in contrast to the scaling of Ref.\cite{CP09}.

Proof that parameter-free scaling is a dynamic scaling  
calls for showing that it connects with stochastic dynamics, i.e., 
it must be shown that Eqs.(\ref{eqx02}) give 
$t_1$ and $t_2$ as functions of $L$ and $p$. 
But Eqs.(\ref{eqx02}) and (\ref{eq07}) are equivalent, 
hence, Eqs.(\ref{eqx02}) do not contain any new  
information above that in Eq.(\ref{eq07}):  
Parameter-free scaling 
does not explain the universal shape of $w(t)$. 
In order to find $t_1$ and $t_2$  one must analyze 
the scale invariance of stochastic growth equation, 
as universal scaling functions express this 
invariance: Parameter-free scaling 
does not express dynamical scale invariance.

Even for models with $p=1$ parameter-free scaling does not 
connect with dynamics. When in Eq.(7) of 
Ref.\cite{CP09} a heuristic parameter $\lambda$ is set $\lambda=1$, 
in order to find $w_2(L)$ and $t_2(L)$ one must perform 
finite-size scaling analysis. On the other hand, when $w_2(L)$ and 
$t_2(L)$ are known beforehand, it is possible to verify if 
Eq.(7) of Ref.\cite{CP09} represents dynamic scaling, which 
shows that there are some examples of  
dynamics where the scaling of Ref.\cite{CP09} may map directly 
on FV function when heuristically $\lambda=1$. However, there 
is no physical reason to set $\lambda=1$. When $\lambda \not=1$, 
then Eq.(7) of Ref.\cite{CP09} does not give FV scaling function 
even if $w_2(L)$ and $t_2(L)$ are known.

Perhaps Eq.(7) in Ref.\cite{CP09} should have been 
supplemented by an explanation that we must always 
have $\lambda=1$ and explicitly know $w_1$, $w_2$, $t_1$, 
and $t_2$  as functions of $L$ and $p$ for a model, in order 
to be able to identify the parameter-free scaling of Ref.\cite{CP09}. 
Of course then $t_1$ and $t_2$ must come from dynamical 
scale-invariance analysis and from FV function, 
because there is no other way to obtain crossover times. 
Therefore the claim made in Ref.\cite{CP09} that 
parameter-free scaling is `something more' than 
FV scaling does not hold: To the contrary, 
parameter-free scaling is `something less.' 
This is because when we know FV function 
we can construct parameter-free scaling 
in such a way that it reflects the scale 
invariance of stochastic dynamics. But if we do 
not know the symmetries of the stochastic growth equation the data 
collapse via parameter-free scaling is meaningless.



\begin{thebibliography}{99}
\bibitem{note01} In RD universality class, 
all curves $w(t)$ for any $L$ follow only one curve 
$w(t)= c \sqrt{t}$, where $c$ is nonuniversal, 
i.e., $c$ is model-dependent and does not 
depend on $L$.
\end{thebibliography}
\end{document}